# A Minimalist Physics-Informed Model for Predicting Extreme Conflict Fatalities

Yair Neuman[1] & Yochai Cohen[2]

[1]. Head, The Functor Lab, Department of Cognitive and Brain Science, Ben-Gurion University of the Negev, Beer-Sheva 84105, Israel,

yneuman@bgu.ac.il

[2]. Gilasio coding, Tel-Aviv, Israel, yohai@gilasio.com

* Corresponding author(s): Yair Neuman (yneuman@bgu.ac.il)

Author's URL of Google Scholar profile:

https://scholar.google.com/citations?user=kHgejLYAAAAJ&hl=en&oi=ao

**Abstract**

The complexity of armed conflicts is expressed in the number of fatalities that may span several orders of magnitude. This study presents a minimalist, physics-informed approach to estimating the likelihood of extreme conflict fatalities at the *country level of analysis* using Bayesian modeling and energy-based dynamics. Leveraging the Boltzmann distribution to construct a Dirichlet prior, we formulate a predictive measure that captures the underlying entropy and energy states of conflict severity. By analyzing a dataset of 112 countries in conflict, we support the predictive power of the proposed measure. The findings suggest that extreme conflict events may be better understood through a minimal but theoretically grounded approach.

**Keywords**: armed conflicts, complex social systems, modeling, Boltzmann, Bayesian

# A Minimalist Computational Model for Extreme Conflict Fatalities

## 1. Introduction

Armed conflicts present a clear signature of a complex system when examined through the perspective of their death toll (i.e., fatalities). The complexity is expressed by the power law distribution of fatalities ([1], [2]). As explained by [3]: "The sizes or severities of wars, commonly measured in battle deaths, have been known since the mid-20th century to follow a right-skewed distribution with a heavy tail, in which the largest wars are many orders of magnitude larger than a 'typical' war."

Drilling down to the *country level of analysis*, which is the *unique* focus of the current study, one may notice that the number of fatalities may span several orders of magnitude. For instance, from 1989 to 2013, Ukraine enjoyed peace. Then, as the war with Russia broke out, the number of Ukrainian fatalities ranged from two orders of magnitude ($N = 254$ fatalities in 2016) to five orders of magnitude in 2022 ($N = 82588$). To illustrate the span, Figure 1 presents the percentage of years per order of magnitude for four countries: Ethiopia (ETH), Rwanda (RWA), Syria (SYR), and Ukraine (UKR).

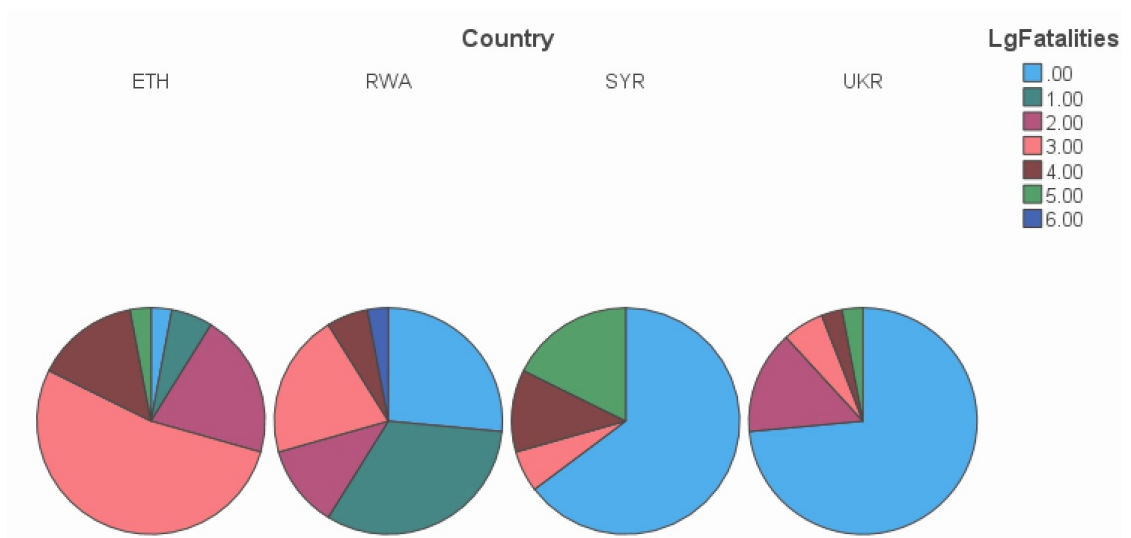

Figure 1. Percent of years per fatality order of magnitude

In this context, where fatalities may span orders of magnitude, an interesting question is whether we can predict the number of years a country experiences a rare number of fatalities according to norms of rarity [4]. More specifically, we ask whether there is a (1) simple and (2) theoretically grounded measure that can support this prediction, regardless of the system's underlying complexity. Here, we use the term prediction in a minimalist sense of model fitting rather than forecasting future events.

The paper is structured as follows. First (section 2.1), we present the pre-processing of a conflict dataset where, for each county, we have the number of fatalities per year. Using a logarithmic transformation, deaths are represented in terms of order of magnitude. For each country, we count the number of years based on the order of magnitude of fatalities. Next (section 2.2), we present the general approach to address the prediction challenge. Section 2.3 details the procedure for computing a simple and theoretically grounded measure we use for prediction. This measure – $P_4$ - is then tested (section 3) in several models predicting the number of years a country will experience a death toll at a rare order of magnitude (dependent variable – $N_4$). Given the success of $P_4$ as an independent variable, we explain it in terms of its ability to update our prior beliefs with respect to the system's energy level and entropy (section 4). We conclude by discussing the ability to understand complex nonlinear systems, such as armed conflicts, using minimalistic models and self-critically point to the shortcomings of the study.

## 2. Methodology

## 2.1. Preprocessing

We utilized the Countries in Conflict Dataset (1989–2022) [5]. The dataset was pre-processed as follows:

1. Fatality Count: For each country and each year, we computed the number of conflict-related fatalities.

2. Binary Classification:

   o If the fatality count was 0 or 1, the country-year was classified as "0" (low or no conflict).

   o Otherwise, the data is transformed as follows:

3. Logarithmic Transformation: The fatality count was transformed using the base-10 logarithm: $L = \text{RND}(\log_{10}S)$, where $S$ is the original fatality count, and RND denotes rounding to the nearest integer.

This procedure maps each year into one of seven orders of magnitude (0-6). We removed from the analysis peaceful countries (e.g., Switzerland) where the distribution of fatalities was concentrated exclusively at level 0. The distribution of years per fatality level is presented in Table 1, with an accompanying rarity scale [4].

| Level i | N | Prob. | Rarity |
|---------|------|--------|-----------------------|
| 0 | 2027 | 0.60 | Not rare |
| 1 | 183 | 0.05 | Moderately rare (R2) |
| 2 | 508 | 0.15 | Frequently rare (R3) |
| 3 | 527 | 0.16 | Frequently rare (R3) |
| 4 | 142 | 0.04 | Very rare (R2) |
| 5 | 12 | 0.003 | Extremely rare (R4) |
| 6 | 1 | 0.0003 | Extremely rare (R4) |

Table 1. Levels of fatalities by rarity scale. Probabilities are rounded.

For the first analysis, we decided to exclude extremely rare fatality levels and focused on predicting $L_4$, which is an order of magnitude defined as "very rare."

## 2.2. The approach

The first approach for predicting the number of fatalities in $L_4$ is to use the number of fatalities in the other levels as predictive features in a model. A reasonable hypothesis is that the higher the number of cases in $L_3$, the more inclined the country is to cross the boundary to $L_4$. We used two measures to test the correlation between the number of cases in $L_4$ (i.e., $N_4$) and the number of cases in the other orders of magnitude: 0, 1, 2, and 3 ($N_0$, $N_1$, $N_2$, and $N_3$, respectively). The Pearson correlation and Spearman's rho are presented in Table 2. We also used a bootstrapping procedure with 1000 to generate the 95% Confidence Interval of the correlations.

| Variable | Spearman's rho | Pearson Correlation |
|---|---|---|
| $N_0$ | -0.56, p< 0.001 CI: -0.67, -0.43 | -0.48, p<0.001 CI: -0.64, -0.39 |
| $N_1$ | ns | ns |
| $N_2$ | ns | ns |
| $N_3$ | 0.60, p<0.001 CI: 0.47, 0.71 | 0.38, p<0.001 CI: 0.22, 0.66 |

Table 2. The correlation between $N_4$ and $N_0$, $N_1$, $N_2$, and $N_3$.

We can see a statistically significant correlation between the number of years classified as $L_4$ (i.e., $N_4$) and the counts in two orders of magnitude: $L_0$ and $L_3$. The more peaceful years a country experiences, the fewer years it experiences with a high number of fatalities. The more years a country experiences at $L_3$, the more years it experiences with a higher number of fatalities at $L_4$.

The counts of years per order of magnitude are based on a *sample* of cases, and therefore, they may have limited *predictive* power for estimating the observed death toll at $L_4$. To address this problem, we used the Dirichlet distribution as a conjugate prior to the multinomial distribution of fatalities in different orders of magnitude. Moreover, we used an energy-based approach to compute this conjugate prior.

Given a distribution of years in different fatality orders of magnitude, we assume that our prior beliefs of observing it can be represented by parameters $\alpha = (\alpha_0, \alpha_1, \alpha_2, \alpha_3)$. These parameters can be used to form the Dirichlet Distribution, which functions as a conjugate prior to the "real" multinomial distribution of fatalities. [6] proposed an energy-based approach to determine the alpha parameters. However, [6] illustrates the approach with respect to a single case only and with respect to changes in the number of fatalities.

The approach trivially assumes that generating higher fatalities requires higher energy levels. Through Boltzmann's seminal work, we know that the probability of a state exponentially and inversely decays with respect to its energy level. This observation is expressed by equation 1, where Z is the normalizing function.

$$P(E_i) = \frac{e^{-E_i/k_BT}}{Z} \qquad (1)$$

Given a distribution of fatalities clustered by their order of magnitude, we may assign energy levels to the different states/classes/levels of fatalities. Here, it must be emphasized that we do not use the Boltzmann distribution in the original sense of macrostates. We use the Boltzmann distribution as an abstract mathematical modeling tool. This is why we describe the approach as physics-informed.

The energy levels we use to model the system have no absolute meaning; they are just relational. Using the Boltzmann distribution, we can generate a probability distribution that includes a state whose probability we would like to estimate (e.g., a year with fatalities in an order of magnitude not yet observed). This distribution is used to compute the *Dirichlet conjugate prior* in order to estimate the posterior multinomial distribution of fatalities by their respective orders of magnitude. In this posterior distribution, our interest is in computing the probability of observing a *single* event (i.e., year) where the number of fatalities is in an order of magnitude, which is defined as a very rare event. In other words, we compute the probability of observing $N_4 = 1$. This probability - $P_4$ - is then used in a model to *estimate* the actual count in $L_4$. The approach is both simple and theoretically grounded. It suggests that the probability of observing a "very rare" event where a country experiences a year with high fatalities is possible when grounded in the (1) observed distribution and (3) the energy level associated with each order of magnitude. The approach is presented and tested in the following sections.

## 2.3. Procedure

Following the transformation of conflict fatalities, we mapped the resulting values into discrete levels, denoted as $L$, where $L \in \{0,1,2,3,4,5,6\}$. First, we focus on levels 0 to 4. Each level corresponds to a distinct classification of conflict severity. We assigned an energy level $E_i$ to each $L_i$ to quantify the severity using a recursive definition based on the midpoint between consecutive levels.

### Step 1. Defining Energy Levels

The midpoint between two consecutive levels of $L_i$ and $L_i+1$ is given by equation 2:

$$M_i = \frac{L_i + L_{i+1}}{2} \qquad (2)$$

which results in values $M_0 = 0.5$, $M_1 = 1.5$, $M_2 = 2.5$, etc.

The energy level for the lowest severity level, $E_0$, is set as $E_0 = 1$.

The energy level for subsequent levels is recursively defined by equation 3:

$$E_i = E_{i-1} + M_{i-1}{}^{\wedge}1.5 \qquad (3)$$

where the exponent 1.5 introduces a nonlinear scaling effect.

Applying this recursive relation, the energy levels are computed as follows:

$E_1 = 1+0.5^{\wedge}1.5 = 1+0.35 = 1.35$

$E_2 = 1.35 + 1.5^{\wedge}1.5 = 1.35 + 1.84 = 3.19$

$E_3 = 3.19 + 2.5^{\wedge}1.5 = 3.19 + 3.95 = 7.14$

and so on.

**Step 2. Computing the Boltzmann distribution**

Next, we followed the approach proposed by [6]. First, we used the $E$ levels computed in step 1 to compute the Boltzmann distribution using equation 1 and $kBT = 1$. The resulting probabilities are presented in Table 3:

| Level i | $P(Ei)$ |
|---------|---------|
| 0 | 0.55 |
| 1 | 0.39 |
| 2 | 0.06 |
| 3 | 0.001 |
| 4 | 0 |

Table 3. The probability of each level of fatalities

Next, we define a sample count vector for each country, where $N_0$, $N_1$, $N_2$, and $N_3$ represent the observed counts (i.e., years) for levels 0-3, respectively. The count for $L_4$ is *always* set to 1. It is important to emphasize that we are interested in predicting the observed number of cases for $L_4$. By heuristically setting the count to 1, we can compute the posterior probability of observing one year in which the country experienced fatalities at level $L_4$. For example, and as presented in Table 4, for Syria, we observe the following distribution of years per fatalities' order of magnitude/level:

| Level i | $Ni$ |
|---------|------|
| 0 | 22 |
| 1 | 0 |
| 2 | 0 |
| 3 | 2 |
| 4 | 4 |

Table 4. Fatalities count per order of magnitude

However, as $N_4$ is what we seek to predict, Table 5 presents a vector where $N_4$ is set to 1:

| Level i | $N_i$ |
|---------|-------|
| 0 | 22 |
| 1 | 0 |
| 2 | 0 |
| 3 | 2 |
| 4 | 1 |

Table 5. Fatalities count with $N_4$ set to "1"

The total count in this case is $s = \sum_{i=0}^{4} N_i = 25$.

**Step 3. Defining the $\alpha$ parameters**

To incorporate prior knowledge, with respect to the probability of observing a state in each energy level, we define the Dirichlet prior α using equation 4:

$$\alpha_i = P(E_i) \cdot s \quad (4)$$

For Syria, this results in the following values presented in Table 6:

| Level i | $P(E_i) \cdot s$ | $\alpha_i$ |
|---------|------------------|------------|
| 0 | 0.55 · 25 | 13.75 |
| 1 | 0.39 · 25 | 9.75 |
| 2 | 0.06 · 25 | 1.5 |
| 3 | 0.001 · 25 | 0.025 |
| 4 | 0 · 25 | 0 |

Table 6. The $\alpha$ parameters for Syria's orders of magnitude

**Step 4. Computing the posterior distribution**

Using Bayes' theorem and equation 5, we compute the posterior distribution of fatalities by updating the prior with the observed counts and the hypothetical count of $L_4$ as:

$$P_i = \frac{N_i + \alpha_i}{s} \quad (5)$$

Where $S$ is computed according to equation 6:

$$S = \sum_{i=0}^{4} (N_i + \alpha_i) \quad (6)$$

Concerning Syria, it resulted in the values presented in Table 7:

| Level i | Sample $N_i$ | Alpha $\alpha_i$ | $Ni + \alpha_i$ |
|---------|--------------|------------------|-----------------|
| 0 | 22 | 13.75 | 35.75 |
| 1 | 0 | 9.75 | 9.75 |
| 2 | 0 | 1.5 | 1.5 |
| 3 | 2 | 0.025 | 2.025 |
| 4 | 1 | 0 | 1 |

Table 7. Posterior counts of observing a year per order of magnitude

The sum of $Ni + \alpha_i$ is $S = 50.025$, which is used to normalize the values to gain the posterior probabilities of observing years in each order of magnitude, as presented in Table 8:

| Level i | Posterior $P_i$ |
|---------|-----------------|
| 0 | 35.75/50.025 = 0.72 |
| 1 | 9.75/50.025 = 0.19 |
| 2 | 1.5/50.025 = 0.03 |
| 3 | 2.025/50.025 = 0.04 |
| 4 | 1/50.025 = 0.02 |

Table 8. Posterior probabilities of observing a year per order of magnitude

Finally, we use the posterior probability for observing a single event at $L_4$ (i.e., $P_4$) and use $P_4$ to estimate the observed number of years in which a country experienced such a magnitude of fatalities (i.e., $N_4$).

## 3. Analysis and Results

### 3.1. Analysis 1.

To test the predictive performance of $P_4$, we first measured the correlation between the observed count in $L_4$ (i.e., $N_4$) and $P_4$. The Pearson correlation and Spearman's rho are presented in Table 9 with their 95% CI.

| Variable | Spearman's rho | Pearson Correlation |
|----------|----------------|---------------------|
| $P_4$ | 0.99, p < 0.001 CI: 0.99, 1 | 0.88, p < 0.001 CI: 0.85, 0.99 |

Table 9. The correlation between $N_4$ and $P_4$.

We can see that $P_4$ is strongly and positively correlated with $N_4$. It outperformed the correlation measured for $N_0$ and $N_3$. This correlation, however, does not tell us about the success in modeling/predicting whether a country will experience a year with a fatality level of $L_4$ (yes or no) and the extent to which we can predict the continuous variable $N_4$ (i.e., the number of years a country experiences fatality level of $L_4$).

A forward linear regression examined the relationship between three independent variables ($N_0$, $N_3$, and $P_4$) and the observed count at $L_4$ ($N_4$). The model was statistically significant, F (2, 109) = 286.808, p < 0.001, explaining 84% of the variance ($R^2 = 0.84$, *Adjusted $R^2$ = 0.837*). The only significant predictors were $P_4$ and $N_3$, with $P_4$ responsible for most of the *R Square Change* (0.767) and $N_3$ adding a relatively small change (0.074). To visually present the relationship between $P_4$ and $N_4$, we first removed the case of Afghanistan, which is an outlier. Next, we transformed the two variables using the Min-max normalization and scaled them to the range of 0-1. Figure 2 presents the relationship between the two variables.

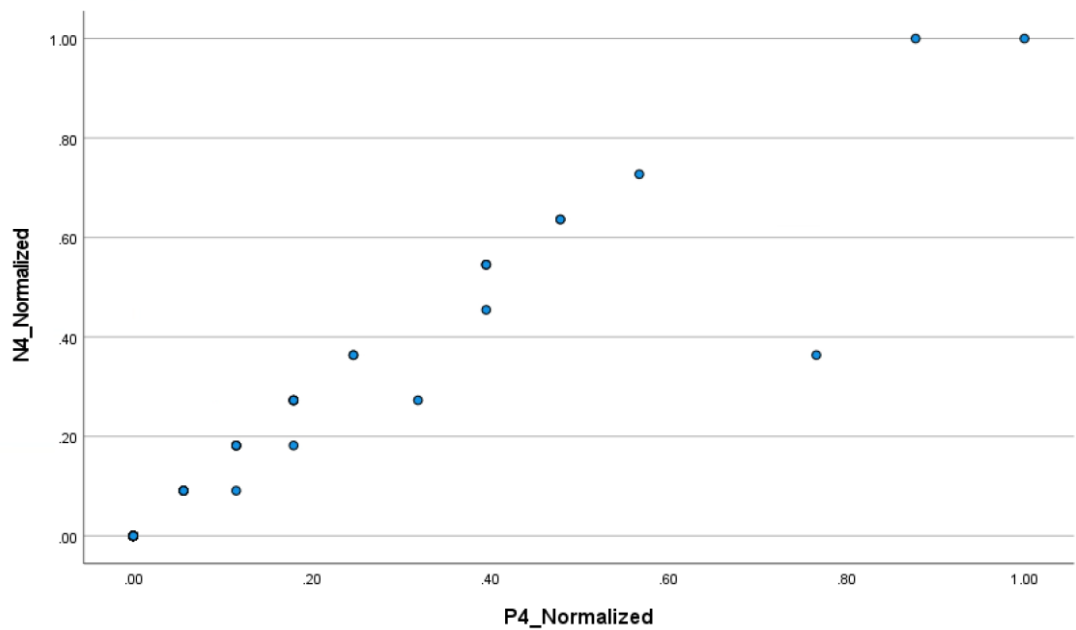

Figure 2. The relationship between $P_4$ and $N_4$

In this case, a linear regression gained F (1, 109) = 1219.037, p < 0.001, explaining 92% of the variance ($R^2 = 0.917$, *Adjusted $R^2$ = 0.918*). We now return to the original data.

We summed the years in the fatality orders of magnitude, which were defined as "rare" (i.e., $\sum_{i=4}^{6} Li$). We examined whether we could predict the number of years classified as rare events for each country (i.e., the dependent variable "rare"). Therefore, we used linear regression with $N_3$ and $P_4$ as independent variables. The model was significant, F (2, 109) = 234.692, p < 0.001, explaining 81% of the variance ($R^2 = 0.812$, *Adjusted $R^2$ = 0.808*).

### 3.2. Analysis 2.

We also tested the predictive power of $P_4$ with respect to $N_4$ using a Neural Network with a *ten-fold cross-validation* procedure. Table 10 presents the average Sum of Squares Error (SSE) and Relative Error (RE) for two models: a model that includes $N_0$, $N_1$, $N_2$, and $N_3$ as independent variables ($M_1$) and a model that includes $P_4$ only ($M_2$):

|        | SSE   |       | RE    |       |
|--------|-------|-------|-------|-------|
|        | $M_1$ | $M_2$ | $M_1$ | $M_2$ |
| Mean   | 6.71  | 0.06  | 0.15  | 0.01  |
| Sd     | 9.72  | 0.03  | 0.12  | 0.007 |
| 95% CI | ± 6.35| ± 0.02| ± 0.08| ± 0.04|

Table 10. Average SSE and RE for the two models

The results clearly show that $P_4$ better predicts the number of fatalities in $L_4$ than the observed counts in the lower orders of magnitude per se.

### 3.3. Analysis 3.

We analyzed $N_4$ as a binary variable ($N_4$_binary) indicating whether a country experienced at least one year with a number of fatalities at the $L_4$ order of magnitude or not (1 vs. 0). A forward logistic regression was conducted to examine the relationship between $P_4$, $N_3$, and the likelihood of experiencing a year or more where the number of fatalities is at the $L_4$ order of magnitude ($N_4 = 0$ or $N_4 = 1$). The logistic regression model was statistically significant, $\chi2(1) = 134.012$, p<.001, with $P_4$ as the only significant variable, indicating that it reliably distinguished between countries that experienced or did not experience fatalities of this order of magnitude. The model explained 100% of the variance (Nagelkerke's $R^2 = 1$).

## 4. Why does it work?

In statistical mechanics, temperature influences the distribution of energy states. In the context of the current study, it affects the spread of probabilities across the states of fatalities. Higher temperatures flatten the distribution, making higher fatality states more probable. Computing the Boltzmann distribution, we generated the probability of observing years with different fatalities orders of magnitude, given the assumed energy level of the state/order of magnitude (i.e., $Ei$) and a temperature $T$ arbitrarily set to "1". However, to understand why $P_4$ is a successful predictor, $T$ can be considered in terms of the ratio between two components: the system's average energy and the system's entropy. The average energy ⟨E⟩ represents the system's general conflict level. It is measured as:

$$\langle E \rangle = \sum p(Ei)Ei \qquad (7)$$

When measuring ⟨E⟩ for each country in our dataset, a higher level of ⟨E⟩ was found to be correlated with the number of years in which a country experienced fatalities at $L_4$ (r = 0.56, p < 0.001) and the sum of years a country experienced fatalities at $L_4$, $L_5$, and $L_6$ (r = 0.57, p < 0.001). The higher the average energy of the system is, the more inclined it is to experience high fatality years. The second component in estimating $T$ is the system's entropy, which, in its most general sense as Shannon entropy, measures the extent to which years are distributed across the orders of magnitude. Given the average energy of the system and its entropy, $T$ can be estimated using equation 8:

$$T \approx \frac{\langle E \rangle}{S} \qquad (8)$$

When computing the Boltzmann distribution to form the Dirichlet prior, we, therefore, set our prior beliefs according to the energy level of each state, the general energy level of the system, and how distributed the years are across orders of magnitude. Here is a concrete example concerning our four-level system composed of four orders of magnitude: $L_0$, $L_1$, $L_2$, and $L_3$. We use the energy levels as explained in section 2.3. For a four-state system with maximal entropy ($S = 2$), meaning the probability of observing each state is equal, the average energy is 3.17, and the estimated $T$ is 1.585. This is an estimation of the system's temperature under *complete ignorance*. Put differently, it is the estimated $T$ of the system assuming *maximal entropy*. When computing Boltzmann with this $T$, representing the system's sensitivity under complete ignorance of the distribution, our degrees of belief reflect the system's "potential" to move to higher energy states (i.e., its temperature) as a function of the system's average energy, and maximally assumed entropy. When recomputing $P_4$ with $T = 1.585$ and using it as the only independent variable in a linear regression model, we found that the model performed better than when T was 1. The model was statistically significant, F (1, 110) = 481.636, p < 0.001, explaining 90% of the variance ($R^2 = 0.814$, *Adjusted $R^2$ = 0.90*). In other words, we explain the success of $P_4$ by interpreting it as (1) The posterior probability of observing cases in $L_4$, (2) A posterior probability, which is grounded in prior beliefs calculated by considering the temperature, (3) a function of the system's energy and (4) entropy. The abovementioned model using $P_4$, as calculated through the average energy and max entropy, both illustrate our explanation and support it by presenting better performance than the model where $T = 1$.

In sum, the minimalistic model presented in this paper involves a measure, which is the outcome of a Bayesian analysis taking into account the observed distribution in different orders of magnitude. However, it is also a measure that integrates the states' energy levels, the distribution's maximal entropy, and the average energy of the observed system.

## 5. Conclusions

Armed conflicts are complex nonlinear systems combining predictable trends with unpredictable behavior (e.g., [7]). In this context, predicting rare events where deaths cross an order of magnitude is important, given our avoidance of thinking about the unthinkable [8]. In this study, we avoided dealing with prediction in its temporal sense and used prediction in a minimalistic sense of model fitting. Therefore, we have no pretension whatsoever to predict when and if a country in conflict will experience a bloody year with a death toll in a rare order of magnitude. Moreover, armed conflicts involve the dynamics of at least two parties. This study ignores this complex interactive aspect and focuses only on model-fitting through the distribution of a single country. Given these qualifications, the study shows that it is possible to estimate the number of years a country experiences fatalities at a high order of magnitude using a Bayesian model that integrates physics-informed measures. The model generates a highly informative measure in various predictive models, from linear regression to neural networks. This measure corresponds with our knowledge of complex systems. As explained [9], the presence of power law distributions in armed conflicts is not an isolated phenomenon; instead, it is observed across various domains, from natural disasters to financial markets. This pattern suggests that war severity may follow the idea of self-organized criticality [10], where, in our context, barriers to high-energy states of fatalities are more likely to be crossed when the average energy level of the system is high. Small perturbations may lead to crossing energy barriers between states and to disproportionately large consequences. Our findings align with this perspective, demonstrating that the probability of a country experiencing a high-magnitude fatality year can be derived from energy-based priors and entropy-informed estimations. Such an interpretation is consistent with prior research emphasizing the role of nonlinear dynamics in war severity [11].

**Authors Contributions**

YN: Conceptualization, research design, statistical analysis, and writing the paper. YC: Data preparation and analysis.

**Funding**

No funding

**Competing interests**

The author(s) declare no competing interests.

**Data availability**

The     dataset     is     available     in     the     following     link:

https://datadryad.org/dataset/doi:10.5061/dryad.z34tmpgrk